\documentclass[letterpaper]{JHEP3}
\usepackage{epsfig}
\usepackage{bbm}
\def\be{\begin{equation}}
\def\ee{\end{equation}}
\def\bea{\begin{eqnarray}}
\def\eea{\end{eqnarray}}
\def\ba{\begin{array}}
\def\ea{\end{array}}

\def\part{\partial}

\def\x{\times}

\def\dag{\dagger}
\def\dagg#1{#1^{\dag}}
\def\bra#1{{\cal{h}} #1 \mid}
\def\ket#1{\mid #1 {\cal{i}}}
\def\ketbra#1#2{\ket{#1} \bra{#2}}
\def\abs#1{\mid #1 \mid}

\def\Id{\ensuremath{\mathbbm{1}}}

\def\vect{\overrightarrow}
\def\eq{\Leftrightarrow}
\def\implique{\Rightarrow}

\def\laplacien{\nabla^2}

\def\qed{\ensuremath{\Box}}

\def\R{\ensuremath{\mathbb{R}}}
\def\C{\ensuremath{\mathbb{C}}}

\newtheorem{theorem}{Theorem}[section]
\newtheorem{lemma}[theorem]{Lemma}
\newtheorem{proposition}[theorem]{Proposition}

\newenvironment{proof}[1][Proof]{\begin{trivlist}
\item[\hskip \labelsep {\bfseries #1}]}{\end{trivlist}}
\newenvironment{definition}[1][Definition]{\begin{trivlist}
\item[\hskip \labelsep {\bfseries #1}]}{\end{trivlist}}

\def\proj{\ket{N-1}\;\bra{N-1} }
\def\comm{\Id - N \proj}

\title{Solitons in finite droplets of noncommutative  Maxwell-Chern-Simons theory}

\author{G. Alexanian\thanks{E-mail: {\tt garnik@lps.umontreal.ca}, present address: Troika-Dialog, 4, Romanov Pereuluk, 125 009 Moscow, Russia}, \  M. Paranjape\thanks{E-mail: {\tt paranj@lps.umontreal.ca}} \\ Groupe de physique des particules, Universit\'e de Montr\'eal \\
C.P. 6128, succ. centre-ville, Montr\'eal, Qu\'ebec, CANADA H3C 3J7 }
\author{I. Pr\'emont-Schwarz\thanks{E-mail: {\tt isabeau.premont.schwarz@gmail.com}, address after September 1st, 2005, DAMTP, Cambridge},\\ Perimeter Institute for Theoretical Physics\\ Waterloo, Ontario, Canada}

\abstract{We find soliton solutions of the noncommutative Maxwell-Chern-Simons theory confined to a finite quantum Hall droplet.  The solitons are exactly as hypothesized in \cite{Manu}.  We also find new variations on these solitons.  We compute their flux and their energies.  The model we consider is directly related to the model proposed by Polychronakos\cite{Poly} and also studied by Hellerman and Van Raamsdonk\cite{HvR}.  They show that this model is equivalent to the Laughlin theory\cite{Laughlin} of the quantum Hall effect.  Our solitons should be thought of as classical vortex configurations of this theory.}


\preprint{UdeM-GPP-TH-05-138}

\dedicated{\`a Daniel Arnaudon, un vrai ami, qui nous a quitt\'e trop t\^ot} 

\begin{document}
\parskip=.5cm
\newpage
\section{Introduction}

It has been know for some time that electrons on a plane in the presence of a strong magnetic field perpendicular to the plane are a realization of a noncommutative geometry.  One such realization was proposed by Susskind\cite{Suss} who describes the quantum Hall fluid by a noncommutative Chern-Simons theory.  Susskind's theory is an effective theory which is lowest  (first) order in derivatives. However it describes only the quantum state of a given conductance plateau.  It does not seem to contain sufficient dynamics to describe transitions between levels, nor the end transition of any given sample to the Hall insulator.   We imagine that such a theory would contain at least two different kinds of excitations, for example plane waves and solitons.  In one region of parameter space the plane waves would be the light and hence dominant mode while in another it would be the solitons.  As a function of a parameter, say the magnetic field, if the two excitations became degenerate, then one would expect a transition in the behaviour of the system at that point.

A pure Chern-Simons theory does not contain excitations.  A possible second order generalization corresponds to adding a Maxwell term to the action.   Such a modified theory contains a richer dynamical content which may allow for a description of the transitions.   A version of noncommutative Maxwell-Chern-Simons theory has been shown to contain vortex-like solitons\cite{KharePar}, however plane wave solutions have not been found there.  In that article the original Chern-Simons action described by Polychronakos\cite{PolyCS1} was used.  This action yields homogeneous equations of motion and it is quite straightforward to find vortex-like solitons once the vacuum configuration has been found.  This version of Chern-Simons theory does not admit a smooth commutative limit.  In a second version of noncommutative Maxwell-Chern-Simons theory, indeed the version studied by Susskind appended by the Maxwell term,  it is possible to find plane wave solutions\cite{Manu}.  However solitons do not seem to exist in this theory.  It seems that the vacuum of the theory is unstable and solitons are possibly incompatible with an infinite plane.   However in the analysis there\cite{Manu}  an intriguing possibility suggested that they might be present in finite quantum Hall droplets.   In this paper we study exactly this theory of finite quantum Hall droplets and confirm the existence of soliton solutions.  We find a rich structure of soliton solutions and we compute their properties and energies.  

\section{Action and equations of motion}

Following Susskind's \cite{Suss} notation we describe the quantum Hall fluid with two fields:
\bea x^i = y^i + \theta \epsilon^{ij}A_j \ \ \ \ \forall i \in \{ 1,2 \} \label{defx} \eea
$A_i $ parametrizes the deviations from the equilibrium situation, $x^i = y^i$, which is a static solution ($\dot{y}^i = 0=\dot{x}^i$ is the fluid velocity) corresponding to the quiescent fluid of uniform density $\rho_0$.   $\theta = \frac{1}{2 \pi \rho_0}$ is the definition of the normalization.
We define the following hermitean scaled covariant derivatives \cite{Manu}:
\bea D_k &=& \sqrt{\theta} (-i \part_{y^k} +  A_k) \ \ \ \forall k \in \{ 1,2 \} \label{defD1} \cr
 D_0 &=& \sqrt{\theta} (-i  \part_t + A_0) \label{defD2} \eea
$A_0$ was introduced by Susskind \cite{Suss} to implement the constraint of the conservation of vorticity of the fluid into the action.
By defining 
\bea D \equiv \frac{D_1 + iD_2}{\sqrt{2}} \ \ , D^{\dag} = \frac{D_1 - iD_2}{\sqrt{2}} \label{defD3} \eea
and supposing that there is a constant magnetic field of strength B perpendicular to the fluid plane with vector potential 
\begin{eqnarray} \vect{\tilde{A}} = 
\left[\begin{array}{c}
\frac{B x^2}{2} \\
\frac{-B x^1}{2} \\
0 
\end{array}\right] 
\label{champmag}
\end{eqnarray}
(one must not confuse $\vect{\tilde{A}}$ with $(A_0,A_1,A_2)$ as the first one describes the real magnetic field, B, and the second one the "position" of the fluid). The action (up to the second order in derivatives) for the fluid including the Maxwell term and the Chern-Simons term can be written as \cite{Manu}:
\begin{eqnarray}
S_{MCS} &=& \frac{(2\pi\theta^{3/2})}{2g^2\,\theta^2}\int d{\tilde t}\;{\rm Tr}\left\{ \left(-2[D_0,D][D_0,D^{\dag}]-[D,D^{\dag}][D,D^{\dag}]\right)+\right. \\ 
& +&\left. {2\lambda}\;\left( -[D,D^{\dag}] + 1 \right)\,D_0\right\} \label{SMCS}
\end{eqnarray}
where $$\tilde{t} \equiv \frac{t}{\sqrt{\theta}}$$ is a rescaled time, $g^2 \equiv \frac{(2 \pi)^2 \rho_0}{m}$, $$\lambda \equiv \frac{eB \theta^{1/2}}{m}$$ and e and m are respectively the charge and the mass of the electron. For simplicity we define: $$\Upsilon \equiv \frac{(2 \pi \theta^{3/2})}{2g^2 \theta^2}$$  The Chern-Simons term taken here corresponds to the one studied by Susskind \cite{Suss}, and admits a smooth commutative limit.  It is for this theory that in \cite{Manu} exact plane wave solutions were found and it was also found that the vacuum solution seems to be unstable.  

In order to reduce our system to a quantum Hall droplet we follow Polychronakos \cite{Poly} and suppose $x^1, x^2$ and $\ A_\mu \ \forall \mu \in \{ 0,1,2 \}$ to be $N \x N$ matrices where $N$ is the number of electrons. Like Polychronakos we subsequently introduce an $N$ dimensional vector $\Psi$ which represents the boundary degrees of freedom.   The boundary term introduced in the action by Polychronakos is $ -2 \Upsilon \ \Psi^{\dag} D_0 \Psi$ which gives the total  action:
\begin{eqnarray}
S &=& \Upsilon \int d{\tilde t}\;\left( {\rm Tr}\left\{
\left(-2[D_0,D][D_0,D^{\dag}]-[D,D^{\dag}][D,D^{\dag}]\right)
\right.\right.\\
&+&\left.\left.  
{2\lambda}\;
\left( -[D,D^{\dag}] + 1 \right)\,D_0\right\} 
- 2 \Psi^{\dag} D_0 \Psi \right)
\label{action}
\end{eqnarray}
Varying with respect to $\Psi^{\dag}$ we get the boundary constraint:
\bea D_0 \Psi = 0 \eq i\dot{\Psi} = A_0 \Psi \label{bndrycstrnt} \eea
Varying with respect to $D_0$ we get the Gauss law:
\begin{eqnarray}
-[D,[D_0, D^{\dag}]]-[D^{\dag},[D_0,D]]-{\lambda}([D, D^{\dag}]-1) -\Psi \Psi^{\dag} = 0
\label{G0}
\end{eqnarray}
Finally, varying with respect to $ D^{\dag} $ we obtain the Amp\`ere law:
\begin{eqnarray} 
[D_0,[D_0,D]]+[D,[D, D^{\dag}]]={\lambda}[D_0,D]
\label{A0}
\end{eqnarray} 
Taking the trace of \ref{G0} gives:
\bea \Psi^{\dag} \Psi = N\lambda \label{norme} \eea
and therefore we can take $\Psi \equiv \sqrt{N\lambda} \ket{N-1}$ where $\ket{N-1}$ is a normalized vector.

\section{The Wigner crystal}

In this section we try to find a ``vacuum" solution, that is a static solution ($\part_t = 0$) with $A_{\mu} = 0$ (and therefore $x^i = y^i$). The fact that $\part_t$ is null imposes that $\ket{N-1}$ is constant. Under those conditions the Gauss law (\ref{G0}) becomes:
\bea [D, \dagg{D}] = \comm \label{G1} \eea
while the Amp\`ere law (\ref{A0}) becomes:
\bea [D, \proj] \equiv D \proj - \proj D = 0 \label{A1} \eea
Replacing for $D$ (with $A_{\mu} = 0$)  we see that equation \ref{G1} is equivalent to:
\bea 
[ \frac{\part}{\part y^1}, \frac{\part}{\part y^2}] = \frac{i}{\theta} \left( \comm \right) \eea
This implies in turn that projecting on the interior of the droplet ($ Vect \{ \ \ \ket{0}, \ket{1}, \ket{2} \cdots \ket{N-2} \ \}$) we have:
\bea [y^1,y^2] = [x^1,x^2] = i \theta \label{noncom} \eea
which endows the droplet with a noncommutative geometry \cite{Connes, Madore}. 

A solution to the Gauss law (\ref{G1}) is the so-called Wigner crystal.  That is:
\bea 
D = a \equiv \sum_{n=0}^{N-2} \sqrt{n+1} \ket{n} \bra{n+1} \ , \label{defa} 
\eea
where $\{ \ \ket{k}   \mid   k \in [ \mid 0, \ N-1 \mid ] \ \}$  is  an  orthonormal  basis. The notation used here and throughout this paper, in general,  ``$k \in [ \mid \alpha, \ \beta \mid ] \ $" signifies that $k$ is an integer that takes values between the integers $\alpha$ and $\beta$ inclusively.
The operators $a$ and $\dagg{a}$ are very similar to the standard annihilation and  creation operators with the notable exception that $\dagg{a} \ket{N-1} = \bra{N-1} a = 0$.
Using \ref{defa}  we have, projecting to the interior of the droplet
\bea 
R^2 \mid_{int.} \equiv {x^1}^2 + {x^2}^2 \mid_{int.}= (\theta /2)(D \dagg{D} + \dagg{D} D) \mid_{int.}=(\theta /2) \sum_{n=0}^{N-2}  (2n+1) \ket{n} \bra{n} \label{rayon} \eea
and so the electrons are spaced out in circles of radius $\propto \sqrt{n}$.  This is the so-called Wigner crystal solution identified by Polychronakos \cite{Poly} and it also corresponds to the ``vacuum" solution of Susskind \cite{Suss}, although he does not mention the connection explicitly.  

In our case of course it is not a solution as it is easy to check that it does not satisfy the Amp\`ere law \ref{A1}.  It was shown that this solution is unstable on the infinite plane \cite{Manu} but now we see that it is (classically) impossible on the droplet with our action.   Actually, under our hypothesis ($\part_t = 0$ and $A_{\mu} = 0$ ) there are no solutions, seeing as \ref{G1} and \ref{A1} are incompatible. In fact, in the ordered base of \ref{defa}, \ref{A1} implies that  
\bea D = \left[ \begin{array}{cc}
{\bf{M}} & \vect{0} \\
\vect{0} & m \end{array} \right] 
\label{matrices} \eea
Where $m \in \C, \ {\bf{M}}$ is an arbitrary $N-1 \x N-1$ matrix and $\vect{0}$ is the null $N-1$ vector. It can easily be seen that this form is incompatible with \ref{G1}.  To find solutions we must therefore modify our hypotheses.

\section{Soliton solutions in the Hall droplet}

\subsection{Arnaudon-Alexanian-Paranjape solution}
Following \cite{Manu} we will look for static solutions $\part_t = 0$ with the following ansatz:

\bea D_0 &=& \sqrt{\theta} A_0 = \sum_{n=0}^{N-1} \phi(n) \ket{n} \bra{n} \cr
D &=& \sum_{n=0}^{N-1} f(n) \ket{n} \bra{n} a  \cr
\Psi &=& \sqrt{N\lambda} \ket{N-1}\label{defDstat} 
\eea 
where  $a$  is  the annihilation operator defined in  \ref{defa} and $\phi (n)$ and $f(n)$ are to be determined by the equations of motion.  The last term of the sum for $D$ is superfluous as it vanishes. We note the following identities which hold for any function $g(n)$, and we will use them in the analysis that follows:
$$ 
[a,(\sum_{n=0}^{N-1}g(n) \ket{n} \bra{n}) ]= a(\sum_{n=0}^{N-1} [g(n)-g(n-1)] \ket{n} \bra{n})= (\sum_{n=0}^{N-1}[g(n+1)-g(n)] \ket{n} \bra{n}) a 
$$

$$ 
[a^\dagger , (\sum_{n=0}^{N-1}g(n) \ket{n} \bra{n})]= a^\dagger (\sum_{n=0}^{N-1}[g(n)-g(n+1)] \ket{n} \bra{n}) =(\sum_{n=0}^{N-1}[g(n-1)-g(n)] \ket{n} \bra{n}) a^\dagger\nonumber 
$$
Notice that $g(-1)$ and $g(N)$ are not defined, although they appear in the equations formally they are not in fact present because they are coefficients to states which are annihilated. 

Let us also calculate the ``magnetic field'', $\tilde B$, and ``electric field'', $\tilde E$, of the fluid:
\bea\nonumber
\tilde B \equiv [D, \dagg{D}]&=&\left( \sum_{n=0}^{N-2}((n+1) \abs{f(n)}^2-n \abs{f(n-1)}^2) \ketbra{n}{n} \right) \\
&-& (N-1) \abs{f(N-2)}^2 \ketbra{N-1}{N-1}
\label{magnetic}
\eea
\be
\tilde E \equiv [D_0, D] =\sum_{n=0}^{N-1}f(n)Q(n) \ketbra{n}{n} a \label{electric}
\ee
where for $  n \in [ \mid 0, \ N-2 \mid ] \ Q(n) \equiv \phi(n) - \phi(n+1)$ and we define  $Q(N-1) \equiv \phi(N-1)$.  Let us also fix $\tilde B(n) \equiv \bra{n} \tilde B \ket{n}$.

The equations of motion with the ansatz \ref{defDstat} gives for the boundary constraint  \ref{bndrycstrnt}:
\bea 0 = i \dot{\Psi} = A_0 \Psi = \sqrt{N\lambda} \phi(N-1) \implique \phi(N-1) = Q(N-1) = 0
\label{cntrnt2}
\eea  
This serves as the boundary condition for the first order difference equation relating $\phi(n)$ to $Q(n)$.  The gauge field equations independently determine the values of the $Q(n)$'s.  The Amp\`ere law \ref{A0} becomes
\bea \sum_{n = 0}^{N-1} f(n) (Q(n)^2 + \tilde B(n+1) - \tilde B(n)) \ketbra{n}{n} a = \lambda \sum_{n=0}^{N-1} f(n) Q(n) \ketbra{n}{n} a \label{A2} \eea
while for the Gauss law \ref{G0} we have:
\bea 
&2& \sum_{n=0}^{N-1} \Big( (n+1) \abs{f(n)}^2 Q(n) - n \abs{f(n-1)} ^2 Q(n-1) \Big) \ketbra{n}{n} \\
&=& \lambda \Big( \tilde B \ - \ \big( \comm \big) \Big) \label{G2} 
\eea
Requiring that $\forall n \in [|0, N-2|], \ f(n) \neq 0$ and replacing for $\tilde B$, \ref{A2} reduces to:
\be 
 Q(n)^2 + (n+2) \abs{f(n+1)}^2 -2(n+1) \abs{f(n)}^2 +n \abs{f(n-1)}^2 = \lambda Q(n) 
\ee
$\forall n \in [| 0, N-3|]$ and 
\be
Q(N-2)^2 -2(N-1) \abs{f(N-2)}^2 +(N-2) \abs{f(N-3)}^2 = \lambda Q(N-2) 
\ee
for $n=N-2$.
Then defining that $f(N-1) = 0$, since $f(N-1)$ does not actually appear in the definition of $D$ we obtain one single defining equation:
\be 
Q(n)^2 + (n+2) \abs{f(n+1)}^2 -2(n+1) \abs{f(n)}^2 +n \abs{f(n-1)}^2 = \lambda Q(n) 
\ee
$\forall n \in [| 0, N-2|]$.
We can rewrite this as:
\be  
\frac{1}{4}(2Q(n)-\lambda)^2 + \laplacien \Big( (n+1) \abs{f(n)}^2 \Big) = \frac{\lambda^2}{4} \label{A2bis} 
\ee
Where $\laplacien h(n) \equiv h(n+1) - 2h(n) + h(n-1)$ is the discrete one dimensional Laplacian.  Furthermore one easily shows \cite{Manu} by induction that \ref{G2} reduces to:
\be  
\left( 2 Q(n) - \lambda  \right) \abs{f(n)}^2 = - \lambda \label{G2bis} 
\ee
$\forall n \in [| 0, N-2|]$.  Then replacing \ref{G2bis} in \ref{A2bis} we get:
\be  
\frac{1}{4} \left( \frac{\lambda}{\abs{f(n)}^2} \right)^2 + \laplacien \Big( (n+1) \abs{f(n)}^2 \Big)= \frac{\lambda^2}{4} 
\ee
$\forall n \in [| 0, N-2|]$.  Finally, defining $u_n \equiv (n+1) \abs{f(n)}^2$ and $g^2 = \frac{\lambda^2}{4}$ we get the following recursion relation:
\begin{equation}
u_{n+1}-2u_{n}+u_{n-1} +g^2\left({n+1}\over u_{n}\right)^2=g^2 \label{recurrence}
\end{equation}
With the boundary conditions that $u_{-1} \equiv  (-1+1) \abs{f(-1)}^2 = 0$ and $ u_{N-1} = N \abs{f(N-1)}^2 = 0$. Which is exactly verifies the suggestion in \cite{Manu}. 
\subsection{New soliton solutions}
Our choice of $\Psi \propto \ket{N-1}$ is quite arbitrary, and one might wonder what would have been different had we chosen $\Psi$ differently. This is what we will now investigate and we will see that it leads to new soliton solutions.  Take 
$$
\Psi = \sum_{n=0}^{N-1} \lambda_n \sqrt{N} \ket{n}.
$$ 
We know from \ref{norme} that
\be 
\sum_{n=0}^{N-1} \abs{ \lambda_n}^2 = \lambda \label{moy2} 
\ee
Keeping the same ansatz (\ref{defDstat}), all our equations remain unchanged until \ref{G2} which becomes:
\bea \nonumber
&2& \sum_{n=0}^{N-1} \Big( (n+1) \abs{f(n)}^2 Q(n) - n \abs{f(n-1)} ^2 Q(n-1) \Big) \ketbra{n}{n} \\
&=& \lambda  \tilde B \ - \lambda \Id + \sum_{m=0}^{N-1} \sum_{n=0}^{N-1} \lambda_m \lambda_n^* \ketbra{m}{n}
\label{G2varpre} 
\eea
Since both the left-hand side of this equation and $\tilde B$ are diagonal we must have  $m \neq n \implique \lambda_m \lambda_n^* = 0$, which in turn implies that $\Psi \propto \ket{M}$ with $M \in [|0,N-1|]$. Thus we take  $\Psi \equiv \sqrt{N \lambda }\ket{M}$ for some fixed but arbitrary value of $M$.  Then the Gauss law (\ref{G2varpre}) becomes:
\bea \nonumber
&2& \sum_{n=0}^{N-1} \Big( (n+1) \abs{f(n)}^2 Q(n) - n \abs{f(n-1)} ^2 Q(n-1) \Big) \ketbra{n}{n} \\
&=& \lambda  \tilde B \ - \lambda \Id + N \lambda \ketbra{M}{M}
\label{G2var} 
\eea 
By \ref{bndrycstrnt} we now have $\phi(M) = 0$ and $\phi(N-1)$ is no longer zero. This is now the modified boundary condition that is used in the difference equation relating the $\phi (n)$'s to the $Q(n)$'s.   One easily shows by induction that equation \ref{G2var} reduces to :
\be 
\left( 2 Q(n) - \lambda  \right) \abs{f(n)}^2 =\cases{ - \lambda &$\forall n \in [|0, M-1|]$\cr
\frac{N \lambda}{n+1} - \lambda& $\forall n \in [|M, N-1|]$} \label{G2varbis} 
\ee
however consistently still imposing $f(N-1)=0$.
Equation \ref{A2bis} is the Amp\`ere law and remains unchanged.  Inserting \ref{G2varbis} in \ref{A2bis} yields:
\be 
\frac{\lambda^2}{4}=\cases{\frac{1}{4} \left( \frac{\lambda}{\abs{f(n)}^2} \right)^2 + \laplacien \Big( (n+1) \abs{f(n)}^2 \Big)&$\forall n \in [| 0, M-1|]$ \cr
\frac{1}{4} \left( \frac{(N-1-n) \lambda}{(n+1) \abs{f(n)}^2} \right)^2 + \laplacien \Big( (n+1) \abs{f(n)}^2 \Big)&$\forall n \in [| M, N-2|]$}
\ee
\newpage\noindent
And so we get the modified recursion relations:
\bea
u_{n+1}-2u_{n}+u_{n-1} +g^2\left({n+1}\over u_{n}\right)^2&=&g^2 \ \ \forall n \in [| 0, M-1|],\cr
 u_{n+1}-2u_{n}+u_{n-1} +g^2\left({N-1-n}\over u_{n}\right)^2&=&g^2 \  \ \forall n \in [| M, N-2|],\label{recurrence2}
\eea
with the same boundary conditions as before,  $u_{-1} = u_{N-1} = 0$.   Thus we have in all $N$ solutions for the ansatz \ref{defDstat}.  We can prove that $\forall g \in \R$ and $\forall M \in [|0,N-1|]$, there exists a unique $u_0 >0$ such that $u_{-1} = 0 = u_{N-1}$ and  $u_n >0 \forall n \in [|1, N-2|]$.  This is intuitively obvious and we relegate a rigorous proof to the appendix.    Furthermore, due to the obvious symmetry of \ref{recurrence2} and the symmetric condition $u_{-1} = u_{N-1} = 0$ we have that: \bea u_{M=m, n=k} = u_{M=(N-1-m), n=(N-2-k)} \label{sym} \eea
One might be lead to think (because of \ref{sym}) that the solutions $M=m$ and $M=N-1-m$ are the same solution with only a permutation of the basis vectors. This is in fact not the case because this permutation (or gauge transformation) does not leave $a$ (as defined in \ref{defa}) invariant.  The energies of these solutions, however, turns out to be degenerate.  Nevertheless, we have $N$ distinct solutions.  The recurrence relations \ref{recurrence2} cannot be solved analytically, in figures 1-4 we show some numerical aspects of the solutions.  

Let remark also that by fixing \bea \Psi = N \lambda e^{\frac{-i \alpha t}{\sqrt{ \theta}}} \ket{M} \label{shift} \eea instead,  we can shift the $\phi(n)$'s by an arbitrary constant $\alpha$.  The boundary condition for the $\phi(n)$'s changes correspondingly.  We can also translate our solutions to the gauge $A_0=0$ albeit time dependent solutions.  Our static solutions relied only on the following commutation relations:
\bea 
[D_0, D] &=&\sum_{n=0}^{N-1}d(n)Q(n) \ketbra{n}{n} a \cr  
[D_0, \dagg{D}] &=&\sum_{n=0}^{N-1} -d(n)^* Q(n) \dagg{a} \ketbra{n}{n} 
\eea
where 
$$
D =\sum_{n=0}^{N-1}d(n) \ketbra{n}{n} a.
$$ 
The second commutation relation is a consequence of the first one if the $Q(n)$'s are real (which they are, according to \ref{G2varbis}). But these commutation relations can be obtained with a time dependent $D$ with $A_0 = 0$ (and thus $D_0 = -i\sqrt{\theta}\part_t $) by taking $d(n) = f(n) e^{i \frac{Q(n)}{\sqrt{\theta}}t}$. The other equations are obviously fulfilled since they depend only on the modulus of $d(n)$ (or $f(n)$).  

\DOUBLEFIGURE{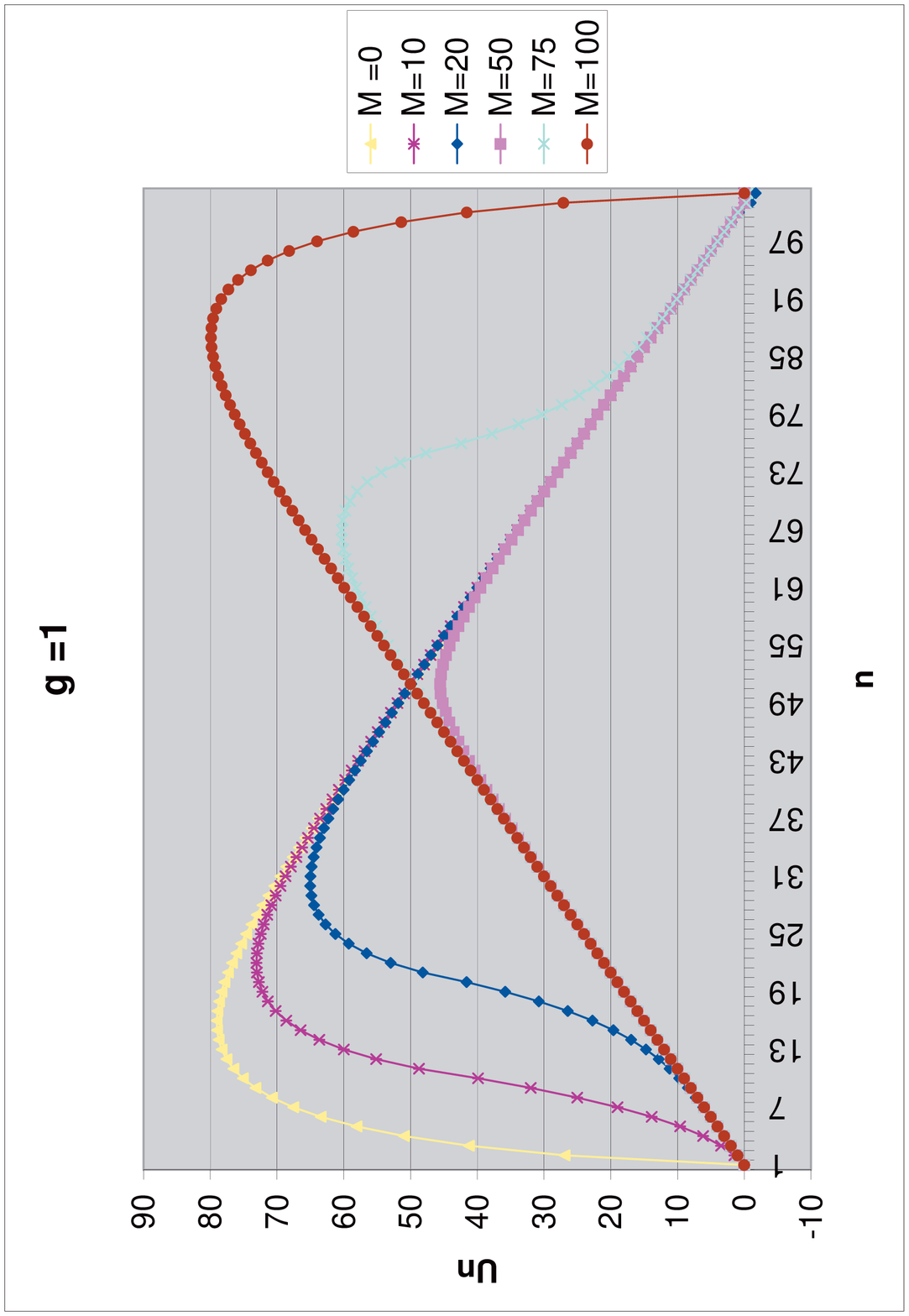,scale=0.3, angle=-90}{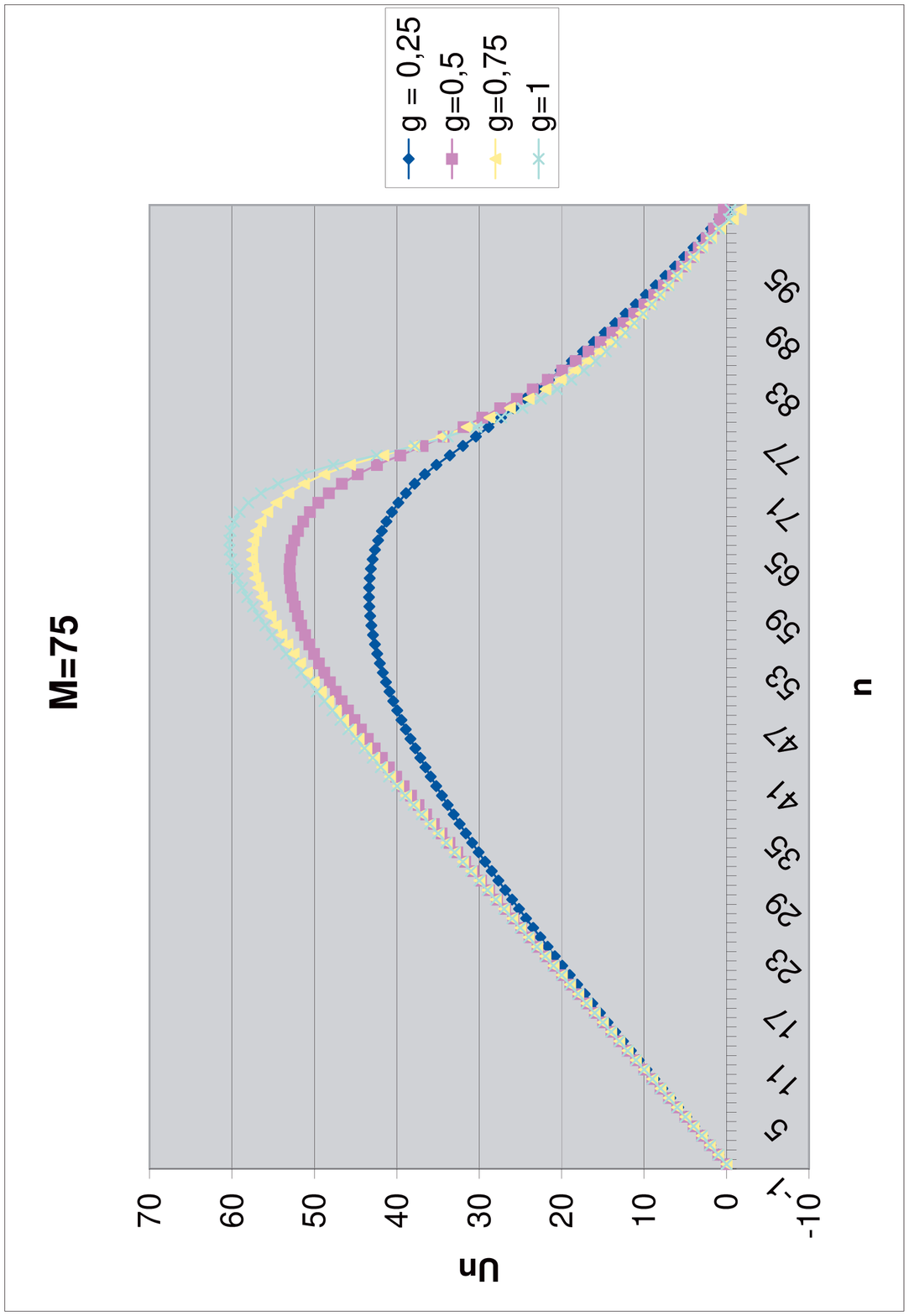,scale=0.3, angle=-90}{Behavior of the solutions of the recursion equations 4.20 for the different values of $M$.  The plot shows $u_n$ for $N = 101$.}{Behavior of the solutions of the recursion equations 4.20 for the different values of $g$. The plot shows $u_n$ for $N = 101$.}


\DOUBLEFIGURE{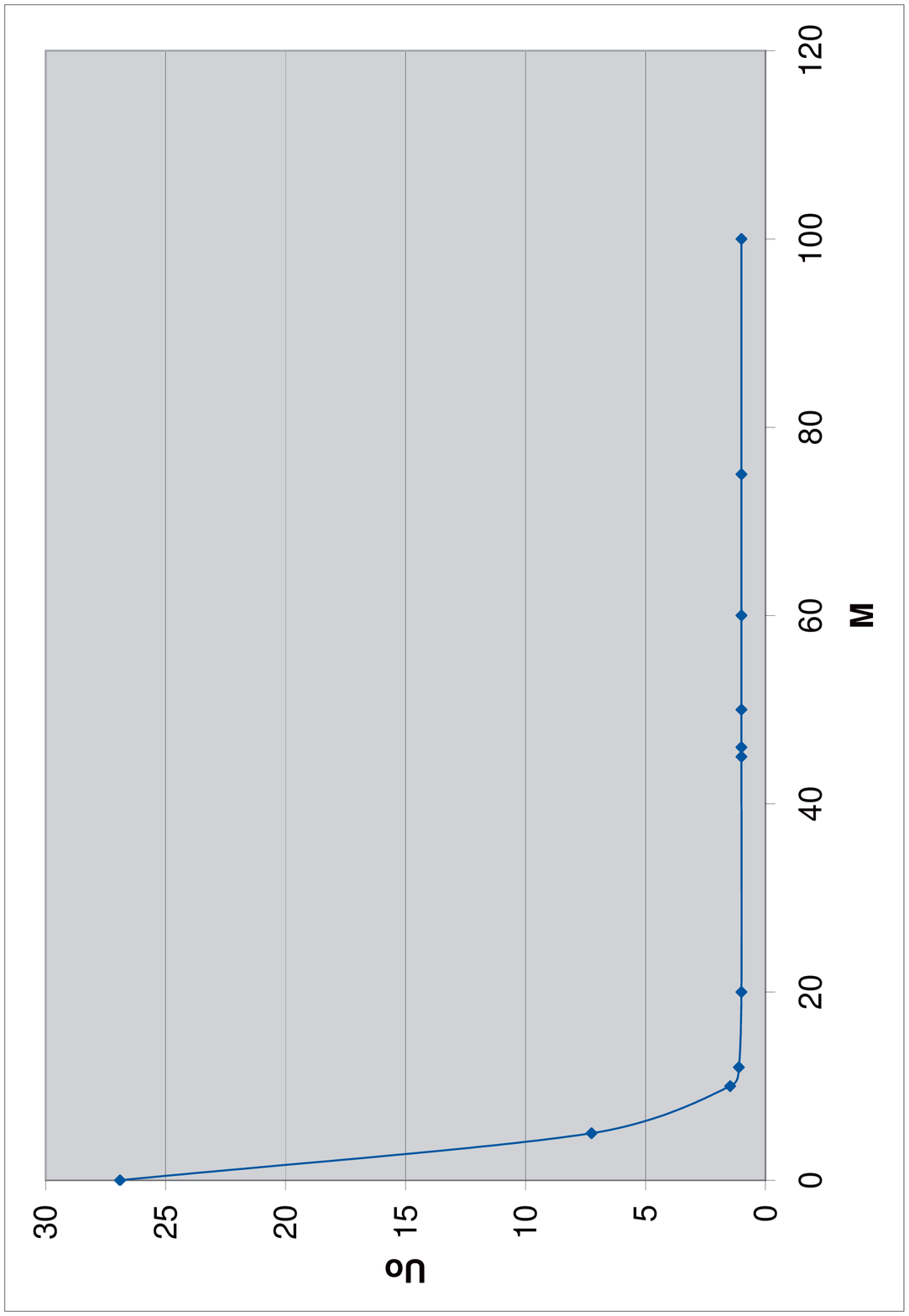, scale=0.3,angle=-90}{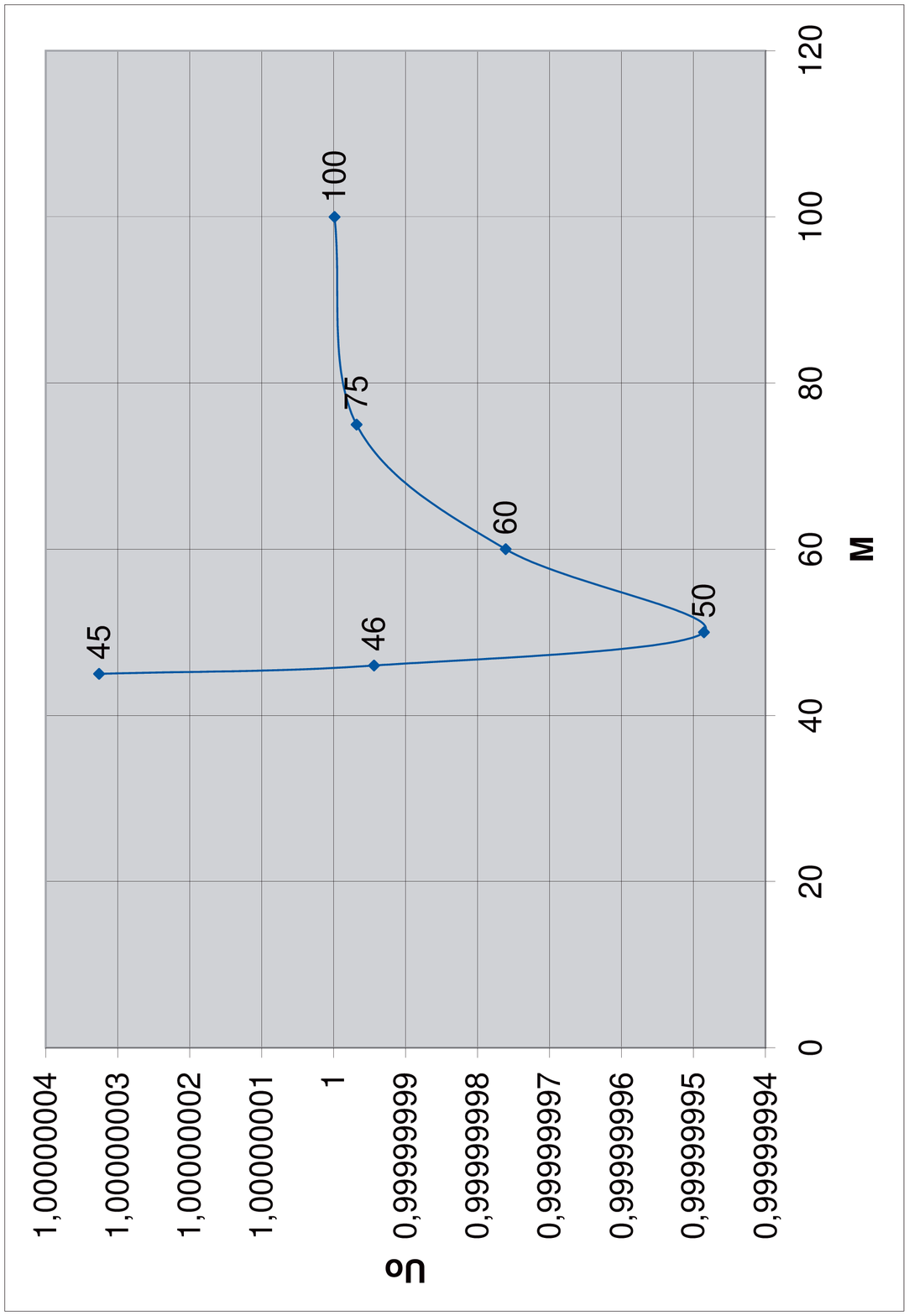, scale=0.3,angle=-90}{Behavior of $u_0$ as a function of $M$. Numerical solutions for $g=1$ and $N=101$.}{Zoom of Figure 3 around the value of $M=50$.}
\newpage
\newpage

\section{Properties of the solitons}
\subsection{Energy}
The Hamiltonian corresponding to action \ref{action} is:
\bea H = \frac{\Upsilon}{\sqrt{\theta}} Tr \Big\{-2[D_0,\dagg{D}][D_0,D]+[D,D^{\dag}][D,D^{\dag}] + D_0 \Big\} \label{Hamiltonien} \eea
The term $Tr D_0 = \sum_{n=0}^{N-1} \phi(n)$ vanishes in the time dependent case and can be put to zero in the static case by adding a phase to $\Psi$ according to \ref{shift}. So, we will henceforth suppose $Tr D_0 = 0$. The kinetic energy is.
\be 
T = \frac{\Upsilon}{\sqrt{\theta}} Tr \Big\{-2[D_0,\dagg{D}][D_0,D] \Big\} = 2 \frac{\Upsilon}{\sqrt{\theta}} Tr \tilde \dagg{E} \tilde E .\label{T} 
\ee
The  potential energy is
\be 
V = \frac{\Upsilon}{\sqrt{\theta}} Tr \Big\{[D,D^{\dag}][D,D^{\dag}] \Big\} = \frac{\Upsilon}{\sqrt{\theta}} Tr \tilde B ^2 .\label{V} 
\ee
For simplicity define $\Xi \equiv \frac{ \Upsilon}{\sqrt{\theta}} = \frac{\pi}{\theta g^2}$ then  replacing for $D$ and $\dagg{D}$ in \ref{T} we get
\be
T = 2 \Xi \sum_{n=0}^{N-2} (n+1) \abs{f(n)}^2 Q(n)^2 = 2 \Xi \sum_{n=0}^{N-2} Q(n)^2 u_n.
\ee
In terms of the $u_n$'s this means:
\be 
T = 2 g^2 \Xi \Big( \sum_{n=0}^{M-1} \frac{\left( u_n - (n+1) \right)^2 }{u_n} \Big) + 2 g^2 \Xi \Big(\sum_{n=M}^{N-2} \frac{\left( N - (n+1) + u_n \right)^2 }{u_n} \Big) \label{T1000}
\ee
Using \ref{recurrence2} we can rewrite this last equation in a form more convenient for numerical analysis:
\be
T =  2 \Xi \Bigg(  \Big\{ \sum_{n=0}^{N-2}  g^2 u_n - u_n \laplacien u_n \Big\} + g^2(M^2+(N-M-1)^2) \Bigg) \label{numérique} 
\ee 
Equation \ref{sym} tells us that the sum in \ref{numérique} is the same for $M=m$ and $M=N-1-m$ and it is easy to verify that the additional constant also has this symmetry. Therefore
\be 
T(M=m) = T(M=N-1-m).\label{symT} 
\ee
Nevertheless, the solutions for $m=M$ and for $M=N-1-m$ are not gauge transforms of one another.  As is clear, the permutation symmetry does not commute with the operator $D$, and the distribution of the ``magnetic" field is different.

Likewise, by replacing for $\tilde B$ in \ref{V},  we get the following expression for the potential energy:
\be 
V = \Xi \sum_{n=0}^{N-1} (u_{n} -u_{n-1})^2  \label{V1000} 
\ee
Due to \ref{sym} we have that:
\be 
V(M=m) = V(M=N-1-m) \label{symV} 
\ee
And so, by adding \ref{numérique} and \ref{V1000} and taking into consideration that $u_{-1} = u_{N-1}=0$ we get the following expression for the total energy:
\be 
E = \Xi \Bigg\{ \sum_{n=0}^{N-2} u_n \laplacien u_n +g^2\Bigg(  \Big\{ 4 \sum_{n=0}^{N-2}  u_n \Big\} + N \Big((N-1)-M \Big) \Bigg)\Bigg\} \label{énergie} 
\ee
In figures 5, 6, 7 we show some numerical aspects of the energy for various values of $M$.  
\DOUBLEFIGURE{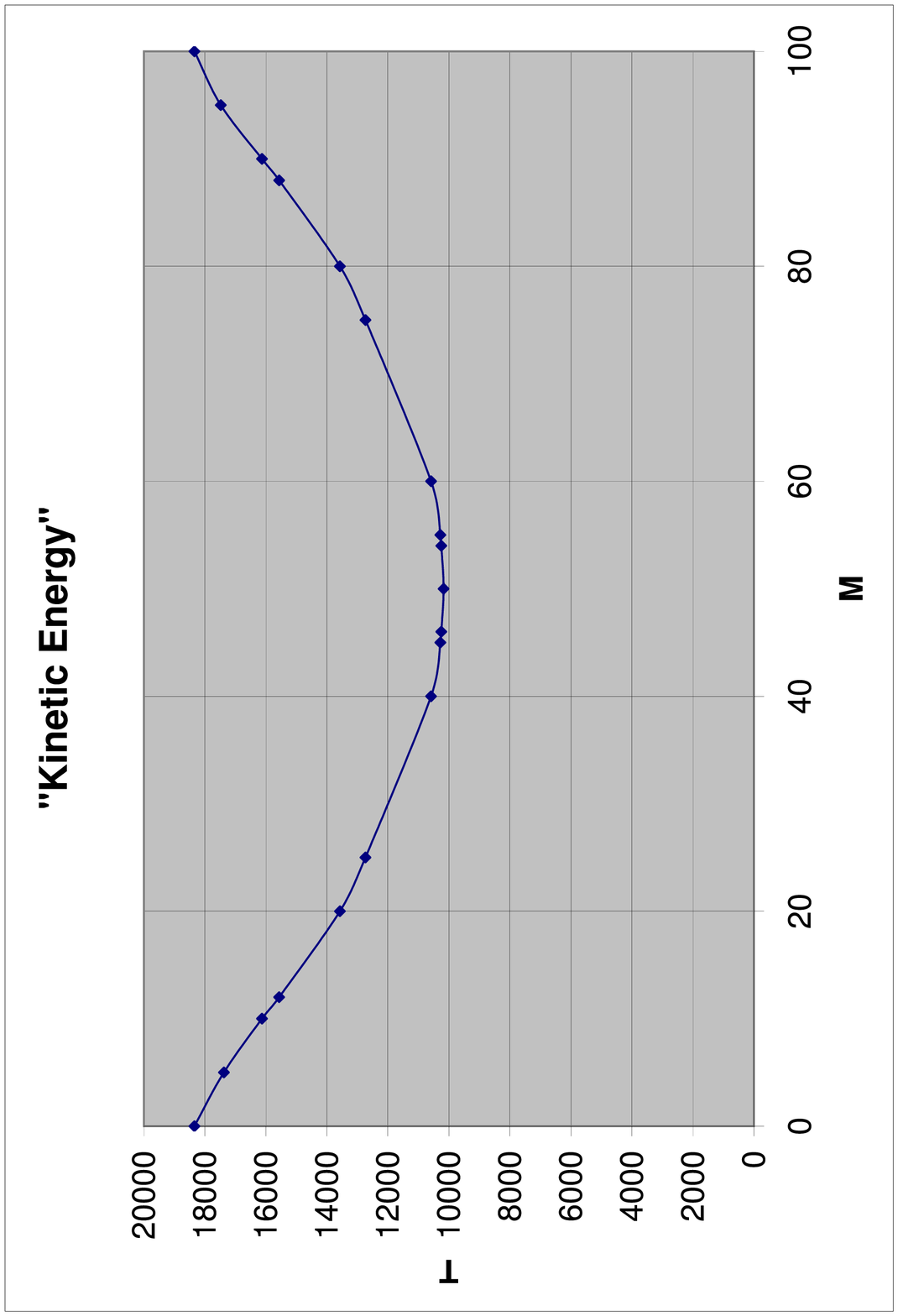, scale=0.3,angle=-90}{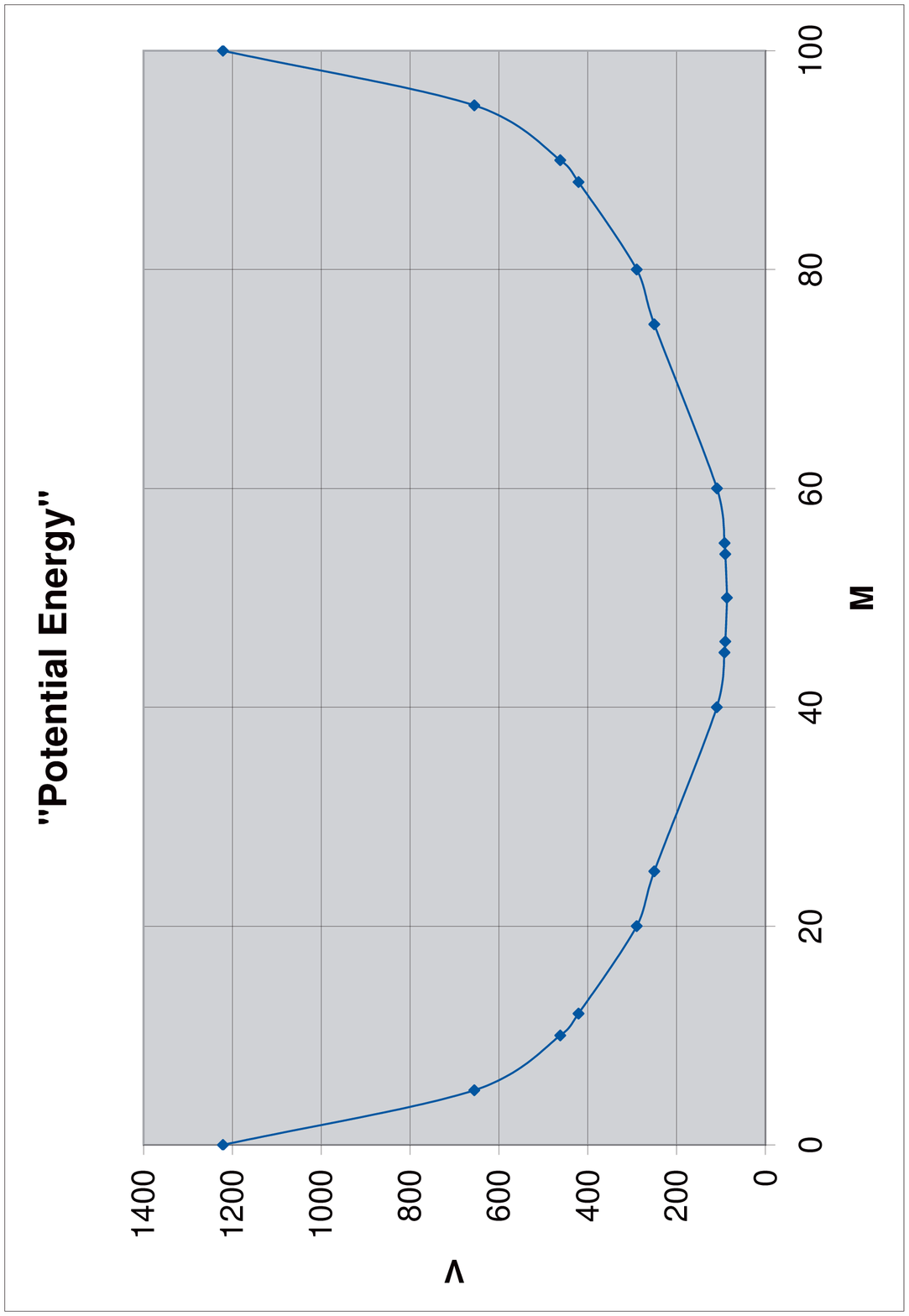, scale=0.3,angle=-90}{Behavior of $T$ as a function of $M$ in units of $\Xi$. Numerical solution for $g=1$ and $N=101$.}{Behavior of $V$ as a function of $M$ in units of $\Xi$. Numerical solution for $g=1$ and $N=101$.}
\FIGURE{\epsfig{file=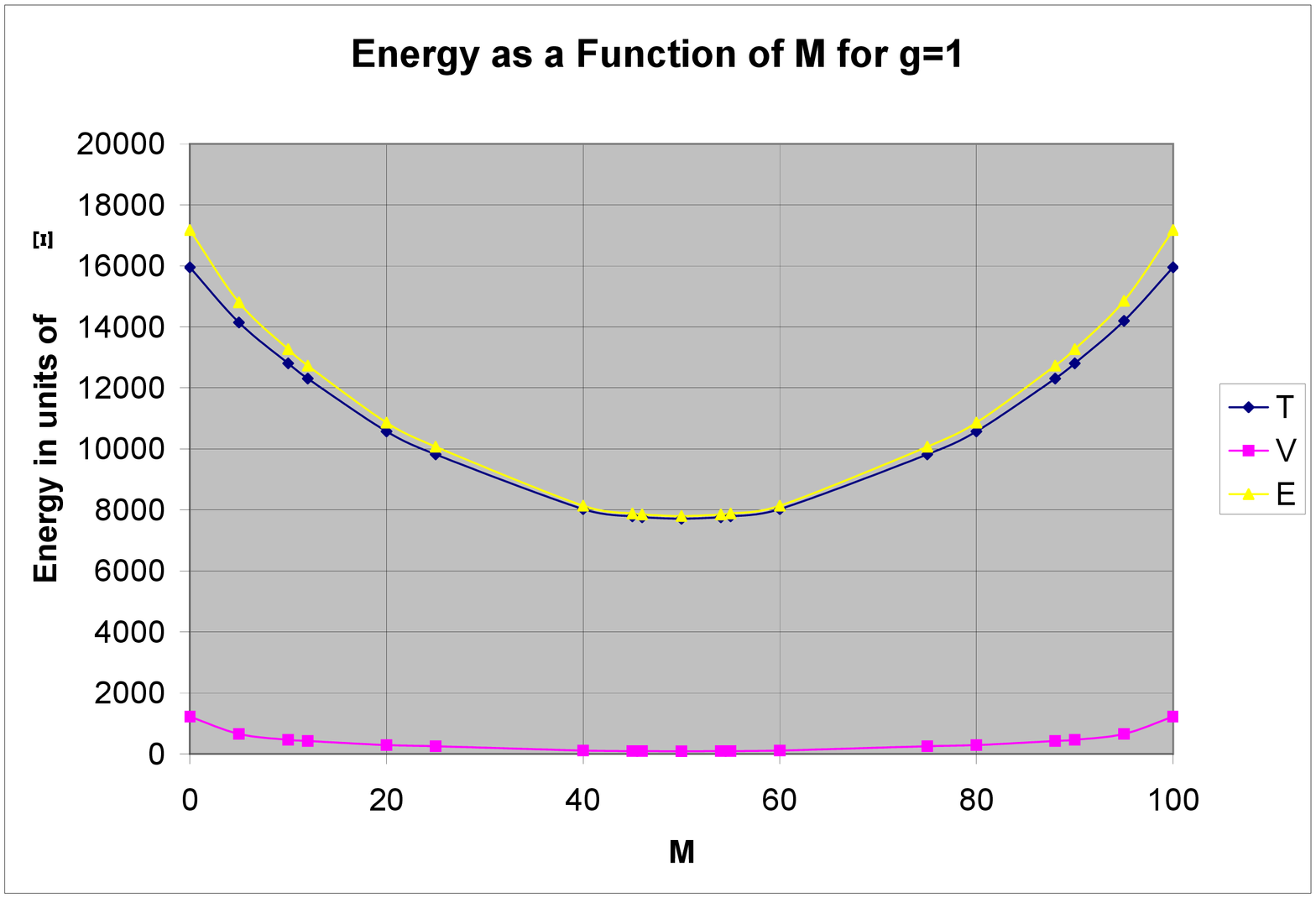,scale=0.6,angle=-90}
\caption{The energy of the solutions for different values of $M$. Numerical values for $g=1$ and $N=101$.}}
\subsection{``Magentic" field and flux}
The total ``magnetic" flux for any of the solitons is strictly zero.  This is a simple consequence of the fact that $\tilde B=[D,D^\dagger ]$ hence the total flux
\be
\Phi_{total} = Tr  \tilde B= Tr [D,D^\dagger ]=0.
\ee
However, if we compute the flux up to a state $\ket {R-1}$ which corresponds to the flux localized in a radius $\sim \sqrt {R-1}$, we find
\bea
\Phi_{R-1} &=& \sum_{m=0}^{R-1}  \bra m\tilde B\ket m= \sum_{m=0}^{R-1}\bra m [D,D^\dagger ]\ket m\\
&=& \sum_{m=0}^{R-1}  \left( u_m-u_{m-1}\right) =u(R-1)=Rf({R-1})^2.  
\eea
Comparing with Figure 1, we see that for different values of $M$, the flux is concentrated over different regions.  For small $M$, there is a positive flux tube at the origin, surrounded by a wide negative flux annular region, while for large $M$ there is a wide cylindrical region of positive flux surrounded by a localized annular region of negative flux near the boundary.  The total flux always vanishes since $u(N-1)=0$.  
\section{Conclusions}
In this paper we have found soliton solutions of the noncommutative Maxwell-Chern-Simons theory, a modified version of the pure noncommutative Chern-Simons theory studied by Susskind \cite{Suss} and which he proposed as the theory of the quantum Hall effect.  In a previous work \cite{Manu} we had studied the modified theory on the infinite noncommutative plane, and we had found plane wave solutions.  Vortex-like soliton solutions were however more elusive and it seems that, in fact, the Susskind vacuum is unstable to the perturbation by the Maxwell term.  In this paper we restrict the Maxwell-Chern-Simons theory to a finite droplet adding boundary degrees of freedom, as proposed by Polychronakos \cite{Poly}.  We find non-trivial soliton like solutions to the equations of motion.  It appears that the energy is minimized for the most symmetric solution which corresponds to a positive ``magnetic" flux cylindrical core over the center half of the droplet surrounded by a negative flux region over the outside half.  It should be noted that this ``magentic" field is an analog of the usual Maxwellian magnetic field and in fact here it actually corresponds to vorticity in the fluid velocity.  The total flux is always zero.  

The model considered is directly related to the model studied by Polychronakos\cite{Poly} and by Hellerman and Van Raamsdonk\cite{HvR}, which is shown to be equivalent to the Laughlin theory\cite{Laughlin} for the quantum Hall effect.  Their model however, contained a harmonic oscillator potential the  sole aim of which was to break the translational invariance of the noncommutative plane.  We do not consider this harmonic potential.  Instead we add the next order corrections, after the Chern-Simons term, in the energy of the velocity of the quantum fluid, the Maxwell term.  We find that adding the Maxwell term a more satisfactory modification of the pure Chern-Simons theory, at least the Maxwell term does measure the actual energy of the flow in the fluid.  With the Maxwell term added we find that there exist multiple classical solutions which are essentially vortices located at the origin superposed with a background of opposite vorticity.  Every physical sample of quantum Hall material exhibits a transition to the Hall insulator at high enough magnetic field\cite{Burgess}.  This transition exhibits a perfect duality in the current to voltage curves just above and just below the transition.  The current flow just above the transition is attributed to charge transport by vortices while just below the transition it is attributed to particles.  The vortices that we have found should model these very physical vortices of the quantum Hall system.  

There are several avenues for future study.  First one should find the analog of plane wave excitations in the model.  Then as a function of the parameters of the model (the external Maxwellian magnetic field, the density (related to the noncommutativity parameter $\theta$ and the coupling constant $g$ (related to the relative strength of the Maxwell term to the Chern-Simons term) it would be very interestiing to find a critical theory where the plane waves and the vortex-like solitons become degenerate.  At this point there should be a phase transition in the behavior of the theory.  This could afford a description of transitions between plateaux or the transition to the Hall insulator that are observed in any experimental Hall system.  Finally, from a more mathematical point of view, it would be interesting to know the full modulli space of solutions to the equations of motion that we have studied.  The stability and other properties of the solutions should be classified.  
\acknowledgments
We thank NSERC of Canada and the London Mathematical Society for financial support.  We also thank Paul Sutcliffe and Camille Boucher-V\'eronneau for useful discussions.  I. P.-S. thanks NSERC for a  University Summer Research Award, 2004, when most of this work was done.  M.P. also thanks Paul Sutcliffe for hospitality at the University of Kent, Canterbury, where this paper was completed and the Kioulafas Scientific Center, Samos for pleasant working conditions, where the final draft was written up.

\appendix
\section{Proof of the existence and the uniqueness of $u_0$ for each $M$}

\suppressfloats
In the following analysis we will assume that $u_{-1} = 0$ and that the $u_n$'s simply satisfy the recurrence relations \ref{recurrence2}.  Then we will prove that there exists a unique value of $u_0$ for which the recurrence relations  \ref{recurrence2}  and the boundary condition $u_{N-1}=0$ are satisfied.
\begin{definition}
Let $\Omega = \{ \ x \in \R^+ \ | \ u_0 = x \implique  u_n > 0 , \ \forall n \in [|0,N-1|]  \}$ 
\end{definition}
\begin{definition}
Let $\tilde \Omega = \{ \ x \in \R^+ | \ u_0 = x \implique u_n > 0 \  \forall n \in [|0,N-2|], \ {\rm and} \ u_{N-1} \geq 0 \}$ 
\end{definition}
\begin{lemma}
\label{nonvide}
$\Omega \neq \emptyset,\ [N, \infty ) \subset \Omega$.
\end{lemma}
\label{lemma1}
\begin{proof}
Take $x \geq N$. We will show that $x\in\Omega$.  We use induction to show that $u_0 = x \implique  u_m \geq N, \ \forall m \in [|0,N-1|],$ and $u_m \geq u_{m-1}$. 
First, if $u_0 = x$, then evidently $u_0 \geq N$ and $u_0 \geq 0 = u_{-1}$.  Thus the induction hypothesis is valid for $u_{-1}$ and $u_0$.  Next we assume that the induction hypothesis is valid for each integer $k$ less than or equal to a fixed $n \in  [|0,N-2|]$. That is, $\forall k \in [|0,n|]$ we assume $u_k \geq u_{k-1} \geq N$.  With this assumption we will prove that $u_{n+1}\geq u_n\geq N$.  By \ref{recurrence2} we have
\be
u_{n+1} = 2u_n - u_{n-1} + g^2\Big( 1 - \frac{{\cal I}(M,n)}{u_n^2} \Big)\label{rec2}
\ee
where 
$$
{\cal I}(M,n)=\cases{(n+1)^2& for $n\in [|0,M-1|]$\cr (N-(n+1))^2& for  $n\in [|M,N-2|]$} .
$$
Evidently, $0< {\cal I}(M,n) <N$.  By the induction hypothesis we have 
$u_n \geq u_{n-1}\geq N$.  Hence  $\frac{{\cal I}(M,n)}{u_n^2} < 1$.  Thus, using the recurrence relation \ref{rec2}, we obtain $u_{n+1} \geq u_n \geq N >0$.  Therefore, by the principle of mathematical induction,  $u_m \geq N > 0,\ \forall m \in [|0,N-1|]$, if 
$u_0 \in [N, \infty)$.   Therefore $[N,\infty) \subset \Omega \neq \emptyset$. \qed
\end{proof}

The $u_n$'s are smooth functions of $u_0$ on $\Omega$, $\forall n \in [|0,N-1|]$, since they never vanish on $\Omega$.  They are in fact simple rational polynomial functions of $u_0$.  Evidently, the only way to introduce a singularity is through the recurrence relations \ref{recurrence2}, which become singular only when any of the $u_n$'s vanish.   This is of course also true on $\tilde\Omega$, since only $u_{N-1}$ may vanish on $\tilde\Omega$, which only introduces a singularity in  $u_N$, which is irrelevant.  Thus $\frac{d \ u_n}{d \ u_0}$ is well defined on $\tilde \Omega$ $\forall n \in [|0,N-1|]$. 
\begin{lemma}
\label{croissant}
$\forall n \in [|0,N-1|], \ \frac{d \ u_n}{d \ u_0} > 0$ and $\frac{d \ u_n}{d \ u_0} > \frac{d \ u_{n-1}}{d \ u_0}$ on $\tilde \Omega$.
\end{lemma}
\begin{proof}
We will again use proof by induction.  First for $u_{-1}$ and $u_0$,  evidently
$\frac{d \ u_0}{d \ u_0} = 1$  and $\frac{d \ u_{-1}}{d \ u_0}=0 $ since $u_{-1}=0$.  Hence $\frac{d \ u_0}{d \ u_0}>\frac{d \ u_{-1}}{d \ u_0}$.
Next we assume that the lemma is true $\forall k \in [|0,n-1|]$ for some fixed $n \in [|0,N-1|]$ and then prove  we that it is true for ${n}$.  This is simply obtained from the recurrence relation \ref{rec2} 
\bea
\nonumber
\frac{d \ u_{n}}{d \ u_0} &=& 2\frac{d \ u_{n-1}}{d \ u_0} - \frac{d \ u_{n-2}}{d \ u_0} + 2 g^2 \frac{{\cal I}(M,n-1)}{u_{n-1}^3} \frac{d \ u_{n-1}}{d \ u_0} \\
&>& \frac{d \ u_{n-1}}{d \ u_0}+ (\frac{d \ u_{n-1}}{d \ u_0} - \frac{d \ u_{n-2}}{d \ u_0}) > \frac{d \ u_{n-1}}{d \ u_0} > 0
\eea
using the induction hypothesis and that ${\cal I}(M,n-1)>0$.  Thus
\be
\frac{d \ u_{n}}{d \ u_0} > \frac{d \ u_{n-1}}{d \ u_0} > 0\ \forall n \in [|0,N-1|]. \ \qed
\ee
\end{proof}

\begin{definition}
Let $\mu = \inf (\Omega)$.   
\end{definition}
$\mu$ is a strictly positive real number since for example, $\Omega \subset (\delta, \infty)$ where  $\delta$ is defined by the value of $u_0=\delta >0$ which renders $u_1=0$,  that is
\be
u_1=0=2\delta+g^2\left(1-\frac{{\cal  I}(M,0)}{\delta^2}\right) .
\ee
Using  our two lemmas, if we start with $u_0>N$ we know we are in $\Omega$, and if we now reduce the value of $u_0$ to $\delta$, we know that the value of $u_1$ will decrease monotonically to zero (we will in fact encounter singularities in the other $u_n$'s already, but for the present purposes these do not matter), hence we are no longer in $\Omega$.  Thus $\Omega \subset (\delta, \infty)$ and $\mu\geq\delta>0$.
\newpage
\begin{proposition}
\label{preuve}
$\mu$ is the unique element of $\R$ such that  $u_0 = \mu \implique  u_n >0, \forall n \in [|1, N-2|]$, and $u_{N-1} = 0$.
\end{proposition}

\begin{proof}
First of all we see that  $\mu = \inf (\Omega) \notin \Omega$.  This is because all of the $u_n$'s are smooth, continuous functions of $u_0$ on $\Omega$.  Thus if $\mu$ were in $\Omega$, then for $u_0=\mu$, $\exists \ \epsilon >0 \ \ni \ u_n>\epsilon \ \forall n \in [|0,N-1|]$.  Hence by continuity of the functions $u_n$ there exists a neighbourhood of $\mu$ for which $u_n>0$ which contradicts the hypothesis that $\mu=\inf (\Omega )$.  This in turn tells us that for $u_0 = \mu$ we have one of the three following possible cases:
\begin{enumerate}
\item 
$\exists n \in [|0,N-1|]$ such that $u_n = 0$ 
\item
$\exists n \in [|0,N-1|]$ such that $u_n < 0$ 
\item
$\exists n \in [|0,N-1|]$ such that $u_n$ is ill defined. 
\end{enumerate}
The third case is clearly a subcase of the first, for if $k$ is the smallest number such that $u_k$ is ill defined.  $k \geq 2$ since $u_0=\mu\ne 0$ and hence $u_1$ is not ill defined, so the first possible ill defined $u_k$ can be $u_2$.  However, by assumption $u_{k-1}$ and $u_{k-2}$ are well defined and then by the recurrence relation \ref{recurrence2}, 
$$
u_{k} = 2u_{k-1} - u_{k-2} + g^2\Big( 1 - \frac{{\cal I}(M,k-1)}{u_{k-1}^2} \Big) .
$$
But this is ill defined only if $u_{k-1} = 0$, thus we are necessarily in the first case.  

The second case can also be easily ruled out.   By continuity of the $u_n$'s, if $u_n(\mu) < 0$ for a sufficiently small $\epsilon > 0$ we still have $ u_n(\mu + \epsilon) < 0$ but $\mu + \epsilon \in \Omega$ as $\mu =\inf (\Omega)$. This is in contradiction with the definition of $\Omega$. 

Therefore only the first case is possible. Suppose now that for some $m \in [|0,N-2|]$ and $u_m(u_0 = \mu) = 0$.  Then by the recurrence relations \ref{recurrence2} 
$$
\lim_{u_0 \to \mu^+} u_{m+1}(u_0) = - \infty .
$$
Then for a finite $\epsilon > 0$, we have $\mu+\epsilon\in\Omega$, but $u_{m+1}(\mu +\epsilon) < 0$ which is a contradiction.  Therefore $u_m(u_0 =\mu) \ne 0 \ \forall m \in [|0,N-2|]$.  

We are therefore forced to conclude that when $u_0 = \mu \implique  u_{N-1} = 0$ and $u_n >0, \ \forall n \in [|1, N-2|]$.

Moreover lemma \ref{croissant} implies that $u_{N-1}$ is an injective function of $u_0$ on $\tilde \Omega$, proving the uniqueness.   Therefore  $\mu$ is the only solution. \qed
\end{proof}


\end{document}